\documentclass[twocolumn,aps,prl,floatfix,superscriptaddress,showpacs]{revtex4}
\usepackage{graphicx}
\usepackage{dcolumn}
\usepackage{bm}
\usepackage{amsfonts}
\usepackage{amssymb}
\usepackage{amsmath}

\usepackage{epsfig}

\DeclareMathAlphabet{\mathitb}{OT1}{cmr}{bx}{sl}

\begin{document}
\pagenumbering{arabic}
  % Title of the article
  \title{Plasmons due to the interplay of Dirac and Schr{\" o}dinger fermions}
  % Authors
  \author{Stefan~Juergens}
  \affiliation{Institute of Theoretical Physics and Astrophysics, University of W\"urzburg, D-97074 W\"urzburg, Germany}
  %\email{michetti@physik.uni-wuerzburg.de, precher@physik.uni-wuerzburg.de}
  \author{Paolo~Michetti}
   \affiliation{Institute of Theoretical Physics and Astrophysics, University of W\"urzburg, D-97074 W\"urzburg, Germany}
  \author{Bj\"orn~Trauzettel}
  \affiliation{Institute of Theoretical Physics and Astrophysics, University of W\"urzburg, D-97074 W\"urzburg, Germany}
  %\pacs{}

  \date{\today}

  \begin{abstract}
    We study the interplay between Dirac and Schr{\" o}dinger fermions in the polarization properties of a two-dimensional electron gas (2DEG).
    Specifically, we analyze the low-energy sector of narrow-gap semiconductors described by a two-band Kane model.
    In the context of quantum spin Hall insulators, particularly, in Hg(Cd)Te quantum wells, this model is named Bernevig-Hughes-Zhang model.
    Interestingly, it describes electrons with intermediate properties between Dirac and Schr{\" o}dinger fermions.
    We calculate the dynamical dielectric function of such a model at zero temperature within random phase approximation.
    Surprisingly, plasmon resonances are found in the intrinsic (undoped) limit, whereas they are absent -- in that limit -- in graphene as well as ordinary 2DEGs.
    Additionally, we demonstrate that the optical conductivity offers a quantitative way to identify the topological phase of Hg(Cd)Te quantum wells from a bulk measurement.
  \end{abstract}

 \maketitle

  %%%%%%%%%%%%%%%%%%%%%%%%%%%%%
  %%%                       %%%
  %%%    INTRODUCTION       %%%
  %%%                       %%%
  %%%%%%%%%%%%%%%%%%%%%%%%%%%%%
  %%%%%%%%%%%%%%%%%%%%%%%%%%%%%%%%%%%%%%%%%%%%%%%%%%%%%%%%%%%%%%%%%%%
  %%%%%%%%%%%%%%%%%%%%%%%%%%%%%%%%%%%%%%%%%%%%%%%%%%%%%%%%%%%%%%%%%%%
  \paragraph{Introduction.}
  The advent of graphene~\cite{geim2007} has paved the way to the investigation of Dirac fermions in condensed matter systems~\cite{neto2009}.
  Since then, Dirac fermion physics has also been analyzed in many other two-dimensional (2D) systems.
  Prominent examples are the surface states of 3D topological insulators (TIs)~\cite{fu2007,hsieh2008},
  where the helicity, i.e. coupling of momentum and spin degrees of freedom, introduces new interesting phenomena~\cite{hasan2010,qi2011}.
  Another rich system is offered by the bulk bands of the 2D TI phase in Hg(Cd)Te quantum wells (QWs)~\cite{konig2007},
  described by the Bernevig-Hughes-Zhang (BHZ) model~\cite{bernevig2006},
  where the interplay of Dirac and Schr{\" o}dinger fermion physics can be studied in combination with topology~\cite{buttner2011}.

  An important aspect of condensed matter physics is the influence of Coulomb
  interaction on observables. In the Dirac fermion system graphene, this research has been intensified in recent years \cite{shung1986, ando2006, barlas2007, hwang2007, wunsch2007, kotov2011}, resulting even in the development of plasmon technology~\cite{grigorenko2012,stauber2013}.
  Plasmons are collective density oscillations commonly occurring at finite doping in an electronic system.
  Most recently, plasmons in Dirac fermion systems have been experimentally observed in graphene~\cite{ju2011, chen2012, fei2012}
  as well as 3D TIs \cite{pietro2013}.
  Usually absent in the intrinsic limit, under certain conditions intrinsic plasmons have been predicted in graphene. For instance, when the electron and hole gas have a finite density due to thermal excitations~\cite{vafek2006,dassarma2013}, or with the inclusion of excitonic effects through ladder-type vertex corrections in the calculation of the dielectric response function~\cite{gangadharaiah2008}. Yet, the latter result is still under debate due to the neglection of diagrams of the same order~\cite{sodemann2012}. Interestingly, a reduction of the dimension to 1D gives rise to intrinsic plasmons in metallic armchair nanoribbons~\cite{brey2007}. These situations are physically distinct from our prediction below where we show that plasmons, in the intrinsic limit, can appear due to an interplay of Dirac and Schr{\" o}dinger physics.

  We present a calculation of the dielectric properties of the BHZ model at zero temperature and doping, within
  random phase approximation (RPA).
  Surprisingly, the interpolating character of the model gives rise to plasmonic resonances, absent in both limiting cases of Dirac and 2DEG system. 
  We discuss their presence and damping rate for varies parameters, including the TI and the normal insulator (NI) phase as well as experimentally realistic values. 
  Furthermore, we calculate the bulk optical conductivity, which offers a way to quantitatively resolve between NI and TI phase.
  %%%%%%%%%%%%%%%%%%%%%%%%%%%%%%%%

  %%%%%%%%%%%%%%%%%%%%%%%%%%%%%%%%%%%%%%%%%%%%%%%%%%%%%%%%%%%%%%%%%%%%%%
  \paragraph{Model.}
  The BHZ model for fermions in Hg(Cd)Te quantum wells -- near the $\Gamma$-point in the Brillouin zone -- has the following form~\cite{bernevig2006}
  \begin{align}
    H = & \left(
      \begin{array}{cc}
	h\left(\boldsymbol{k}\right) & 0\\
	0 & h^{*}\left(\boldsymbol{-k}\right)
      \end{array}\right),
   \label{eq:Hamiltonian}\\
    h\left(\boldsymbol{k}\right) =& V(k) + \boldsymbol{d}_{\boldsymbol{k}}\cdot\vec{\sigma},\nonumber \\
    \boldsymbol{d}_{\boldsymbol{k}} =& \left(\begin{array}{ccc}
    Ak_{x}, & Ak_{y}, & M\left(k\right)\end{array}\right), \nonumber
  \end{align}
  where $\vec{\sigma}$ are the Pauli matrices associated with the
  band-pseudospin degree of freedom (subband $E_{1}$ or $H_{1}$ in HgTe QWs), $V(k)=C-Dk^{2}$ and $M(k)=M-Bk^{2}$.
  The system possesses time-reversal symmetry and $H$ is block diagonal
  in the Kramer's partner or spin degree of freedom.
  Therefore, we restrict ourselves to the block $h\left(\boldsymbol{k}\right)$,
  from which the results can be extended to the other one by applying the time reversal operator.
  %$\hat{T}=is_{y}\hat{K}$.

\begin{figure}
    \includegraphics[width=4.2cm]{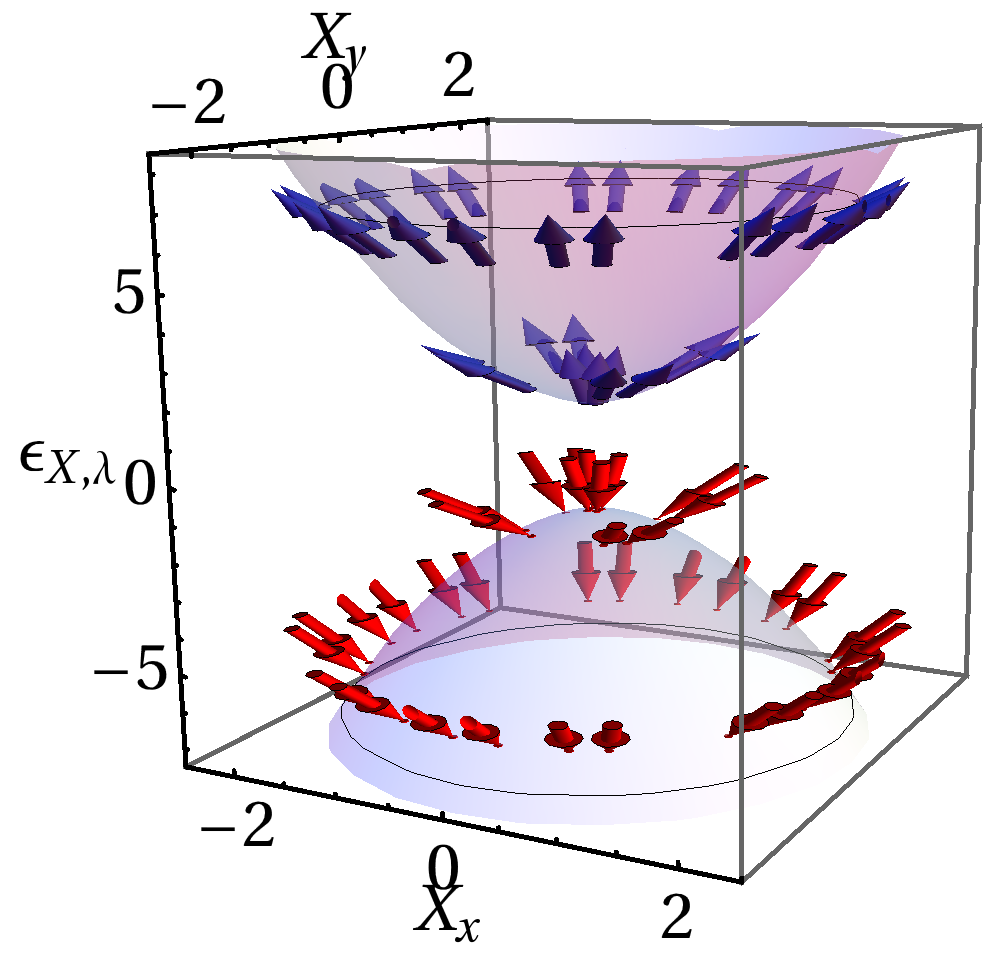}
    \includegraphics[width=4.2cm]{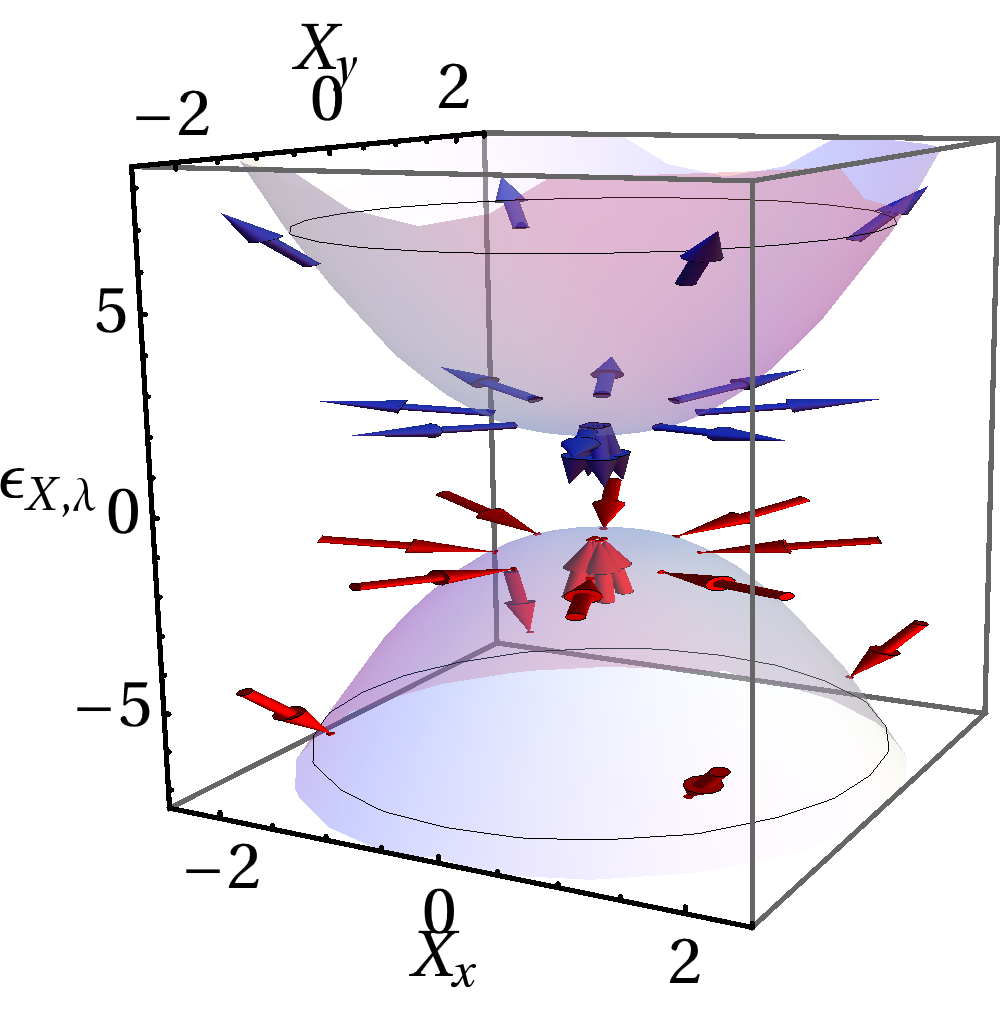}
    \caption{Dispersion relation and pseudospin properties of the eigenstates of the BHZ model for
    a NI phase with $\xi_M=\frac{1}{2}$ (a) and a TI phase with $\xi_M=-\frac{1}{2}$ (b).
    The two bands have been artificially seperated by an additional $2 \epsilon_{X,\lambda}$ for a better illustration of the pseudospin.
    }
  \label{fig:pseudospin}
  \end{figure}

  Evidently, $h\left(\boldsymbol{k}\right)$ smoothly interpolates, as a function of its parameters,
  between a Dirac and a conventional 2DEG system.
  The off-diagonal term ($A$ parameter) is typical of
  a Dirac system with $M$ the Dirac mass.
  Negative masses correspond to the TI phase described by a finite $\mathbb Z_2$ topological invariant~\cite{kane2005},
  while positive masses lead to a normal insulator (assuming $B<0$).
  In analogy to a 2DEG, the diagonal parts bear kinetic energy elements which preserve ($B$ parameter)
  and break ($D$ parameter) particle-hole symmetry. %~\cite{D}.
  The eigenstates of Eq.~(\ref{eq:Hamiltonian}) are characterized by energy dispersion and pseudospin~\cite{pseudo}:
  \begin{eqnarray}
    E_{k,\lambda} = V(k) + \lambda~|\boldsymbol{d}_{\boldsymbol{k}}| \; ,
    \label{eq:Dispersion}\\
    \langle\boldsymbol{k}, \lambda|~\vec{\sigma}~|\boldsymbol{k}, \lambda\rangle = \lambda \hat{\boldsymbol{d}}_{\boldsymbol{k}}
  \end{eqnarray}
  with $\lambda=\pm$ for conduction and valence bands, where $\hat{\boldsymbol{d}}$ is the direction of the $\boldsymbol{d}$ vector,
  illustrated in Fig.~\ref{fig:pseudospin}.

  %%%%%%%%%%%%%%%%%%%%%%%%%%%%%%%%%%%%%%%%%%%%%%%%%%%%%%%%%%%%%%%%%%%%%%%%%%%%%%%%%%%%%%
  %%%%%%%%%%%%%%%%%%%%%%%%%%%%%%%%%%%%%%%%%%%%%%%%%%%%%%%%%%%%%%%%%%%%%%%%%%%%%%%%%%%%%%
  \paragraph{Polarization function.}
  At zero doping, there are no extrinsic parameters like the Fermi momentum.
  Nevertheless, the parameters of the BHZ model provide us natural units for the wave vector $q_0=A/|B|$ and the energy $E_0=A q_0$. Experimentally relevant values for $q_0$ and $E_0$ are discussed at the end of the plasmon section. 
  We define the dimensionless wave vector $\boldsymbol X=\boldsymbol q/q_0$ and frequency $\Omega= \hbar \omega/E_{0}$,
  as well as the dimensionless parameters $\xi_{M}=M/E_{0}$ and $\xi_{D}=D/|B|$. 
%  %
%  Using these natural scales, this choice reduces the parameter space considerably, e.g. the eigenenergy $E_{X,\lambda}=E_{0}\epsilon_{X,\lambda}$
%  is linearly scaling in $E_{0}$ and only a function of $X$ and $\xi_{M}$.
%  %
  Hence, the physical scales of momentum and energy are fixed by the ratio of the $A$ parameter (linked to the Dirac physics) to
  the $B$ parameter (linked to the 2DEG physics).
  In terms of the dimensionless variables,
  we expect, therefore, an intermediate behavior for $X\sim1$, while for
  $X,\Omega\rightarrow0$ ($X,\Omega\rightarrow\infty$) the Dirac (2DEG) physics should emerge.

  The linear response of a homogeneous system to an external applied
  potential is described by the polarization function
  $\Pi^{R}\left(\boldsymbol{q},\omega\right)$.
  This response comprises the two main phenomena of screening and dissipation,
  included in the real and imaginary part of $\Pi^{R}\left(\boldsymbol{q},\omega\right)$, respectively.
  The intrinsic polarization function of the BHZ model in RPA approximation at zero temperature,
  where only interband terms contribute, can be written as
  \begin{equation}
    \Pi^{R}\left(X,\Omega\right) = \frac{g_{s}}{\left|B\right|}\hspace{-0.05cm}
    \underset{\lambda=\pm}{\sum}\hspace{-0.05cm}
    \int\hspace{-0.15cm}\frac{d^{2}\hspace{-0.05cm}\tilde{X}}{\left(2\pi\right)^{2}}\hspace{0.05cm}
    \frac{\lambda~\mathcal{F}\left(\boldsymbol{\tilde{X}},\boldsymbol{\tilde{X}'}\right)}
    {\Omega\hspace{-0.05cm}+\hspace{-0.05cm}i{\rm 0^{+}\hspace{-0.05cm}+\hspace{-0.05cm}
     \epsilon_{\tilde{X}\hspace{-0.01cm},\hspace{-0.01cm}-\hspace{-0.03cm}\lambda}\hspace{-0.1cm}-\hspace{-0.03cm}\epsilon_{\tilde{X}',\lambda}}}
    \label{eq:Pi_par}
  \end{equation}
  with $\boldsymbol{\tilde{X}'}=\boldsymbol{\tilde{X}}+\boldsymbol{X}$, $0^{+}$ a positive infinitesimal, $g_{s}=2$ for spin degeneracy, $\epsilon_{\tilde{X},\lambda}=E_{q_0 \tilde{X},\lambda}/E_0$
  the dimensionless eigenenergies, and
  \begin{equation}
    \mathcal{F}\left(\boldsymbol{X},\boldsymbol{X'}\right)=
    \left|\langle \boldsymbol k,+|\boldsymbol{k'},-\rangle\right|^2=
    \frac{1}{2}\left[1-\hat{\boldsymbol{d}}_{q_0\boldsymbol{X}}\cdot\hat{\boldsymbol{d}}_{q_0 \boldsymbol{X'}}\right].
    \label{eq:F Factor}
  \end{equation}
  From Eq.~($\ref{eq:Pi_par}$), we see that $\left|B\right|\Pi^{R}\left(X,\Omega\right)$
  is a function of $X$ and
  $\Omega$, while it parametrically depends on $\xi_{M}$ and $\xi_{D}$.

  The same is also true for the dynamical dielectric function, which acquires the form
  \begin{equation}
    \frac{\varepsilon\left(X,\Omega\right)}{\varepsilon_{r}}= 1 -\alpha~g\left(X,\Omega\right),
    \label{eq:epsi}
  \end{equation}
  where we define the effective fine structure constant
  $\alpha=\frac{1}{A}\frac{e^{2}}{4\pi\varepsilon_{0}\varepsilon_{r}}$~\cite{kotov2011}.
  In graphene, it is of the order $\alpha=2.2/\varepsilon_{r}$~\cite{grigorenko2012}; in HgTe, one finds $\alpha\approx4/\varepsilon_{r}$~\cite{buttner2011,schmidt2009}.
  Here, $\varepsilon_{r}$ is the background dielectric constant and we introduce
  the function $g\left(X,\Omega\right)=2\pi\frac{\left|B\right|}{X}\Pi^{R}\left(X,\Omega\right)$.
  While $\varepsilon_{r}$ accounts for the screening of internal electronic shells, $-\alpha g\left(X,\Omega\right)$
  yields the dynamical screening due to valence electrons within RPA.

  Beside the screening properties, the dielectric function also provides insight into the excitations of the system.
  Zeros of $\varepsilon\left(X,\Omega\right)$ identify momentum and frequency at which the system
  allows for self-sustaining density perturbations, which form collective excited modes of the system, i.e. plasmons.
  The dispersion relation of the plasmonic solutions is obtained by solving~\cite{Fetter, giuliani}
  \begin{equation}
    \varepsilon\left(X,\Omega_{p}-i\Gamma\right)=0,
    \label{eq:Plasmon Equation}
  \end{equation}
  where $\Omega_{p}$ is the dimensionless eigenfrequency of the plasmon and the finite imaginary
  part $\Gamma=\frac{\gamma}{E_{0}}$ accounts for the possible damping due to single-particle excitations.

  Dissipation processes in the interacting system are described by $\varepsilon$ through the loss function
  \begin{equation}
    \Im\left[-\frac{1}{\varepsilon}\right]=\frac{1}{\varepsilon_{r}\alpha}
    \frac{\Im\left[g\left(X,\Omega\right)\right]}{\left|\frac{1}{\alpha}-g\left(X,\Omega\right)\right|^{2}},
    \label{eq:Loss function}
  \end{equation}
  which accounts for the excitation of both single-particle electron-hole pairs and collective plasmon modes.
  %

  %Surprisingly, Eqs.~(\ref{eq:Pi_par},~\ref{eq:epsi},~\ref{eq:Loss function}) show that just the parameters $\xi_{M}$, $\xi_{D}$ and $\alpha$ determine the complete
  %intrinsic physics of the BHZ model including Coulomb interaction within RPA.
  %
 % This enables us, in the following, to draw a very general picture of the different physical aspects, including overlap factor, polarization function,
  %plasmons, and optical conductivity.

  %%%%%%%%%%%%%%%%%%%%%%%%%%%%%%%%%%%%%%%%%%%%%%%%%%%%%%%%%%%%%%%%%%%%%%%%%%%
  %%%%%%%%%%%%%%%%%%%%%%%%%%%%%%%%%%%%%%%%%%%%%%%%%%%%%%%%%%%%%%%%%%%%%%%%%%%

  \paragraph{Overlap factor.}
  %We begin our discussion with the differences in the overlap factor $\mathcal{F}$, arising from the pseudospin
  %structure of the eigenstates as depicted in Fig.~\ref{fig:pseudospin} for a NI phase (a) and TI phase (b), compared to the Dirac limit.

  Massless Dirac fermions (corresponding to $D=B=M=0$) are characterized by their helicity and consequently the overlap factor reduces to
  $\mathcal{F}(\boldsymbol{k},\boldsymbol{k'})=\frac{1}{2}\left(1-\cos{\theta}\right)$,
  with $\theta$ the angle between $\boldsymbol{k}$ and $\boldsymbol{k'}$.
  Hence, it is strictly one (zero) for $\boldsymbol{k}$ and $\boldsymbol{k'}$ pointing in the opposite (same) direction.
  Eigenstates of the BHZ model are characterized by their pseudospin.
  The quadratic terms turn the pseudospin out of plane in opposite directions
  for conduction and valence bands at large $X$, see Fig.~\ref{fig:pseudospin}.
  This behavior results in a decay of the overlap factor for $A\rightarrow0$ or $X\rightarrow\text{\ensuremath{\infty}}$,
  reproducing the limit of a conventional 2DEG system where intrinsic polarization is absent.
  For $X\ll|\xi_M|$, a finite Dirac mass term turns the pseudospin out of plane as well, in the same direction as the quadratic terms [Fig.~\ref{fig:pseudospin}(a)]
  for $M>0$ (NI phase), or in the opposite direction [Fig.~\ref{fig:pseudospin}(b)] for $M<0$ (TI phase). This behavior suppresses the overlap factor.
  In the TI phase, valence and conduction band states with unitary overlap can always be found for finite $X$. Thus, the overlap factor is enhanced in the region $X\gtrsim|\xi_M|$ with respect to the NI phase. These observations will help us to better understand the subsequent physics.

  %%%%%%%%%%%%%%%%%%%%%%%%%%%%%%%%%%%%%%%%%%%%%%%%%%%%%%%%%%%%%%%%%%%%%%%%%%%
  %%%%%%%%%%%%%%%%%%%%%%%%%%%%%%%%%%%%%%%%%%%%%%%%%%%%%%%%%%%%%%%%%%%%%%%%%%%
  \paragraph{Results.}

\begin{figure}
    \includegraphics[width=4.2cm]{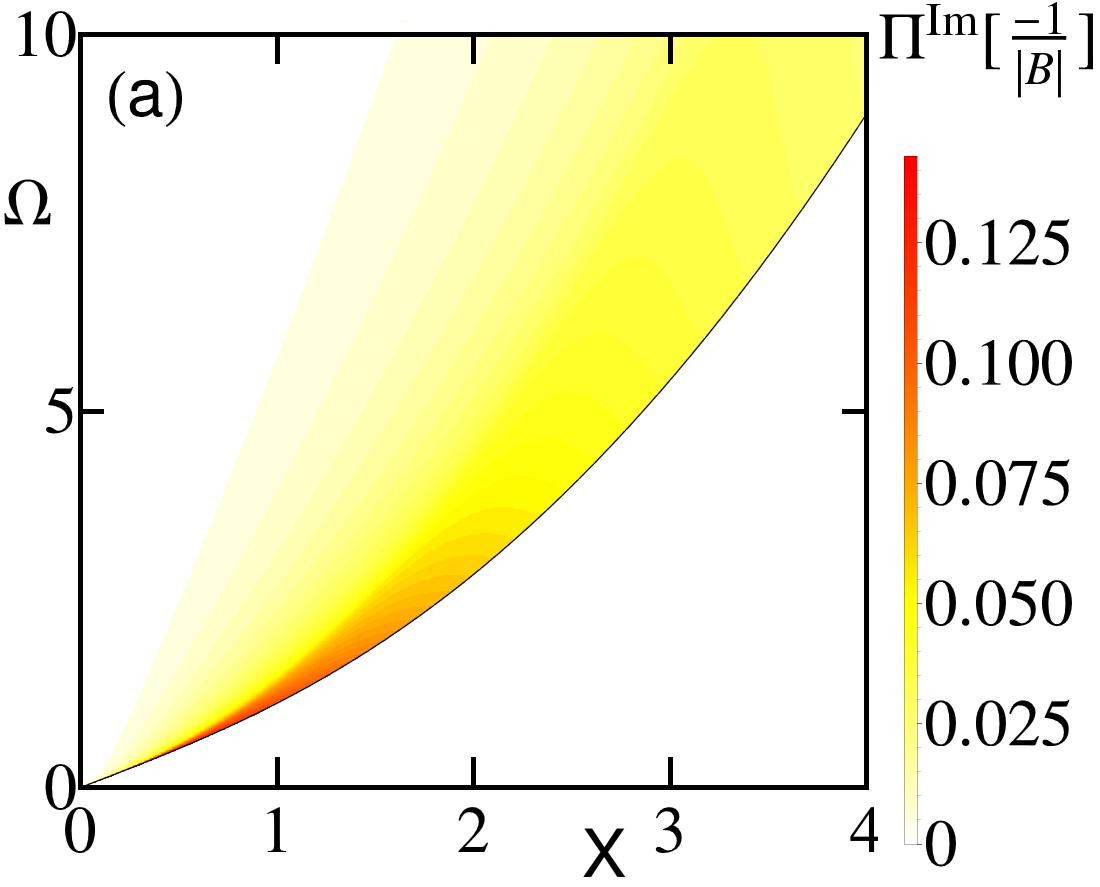}
    \includegraphics[width=4.2cm]{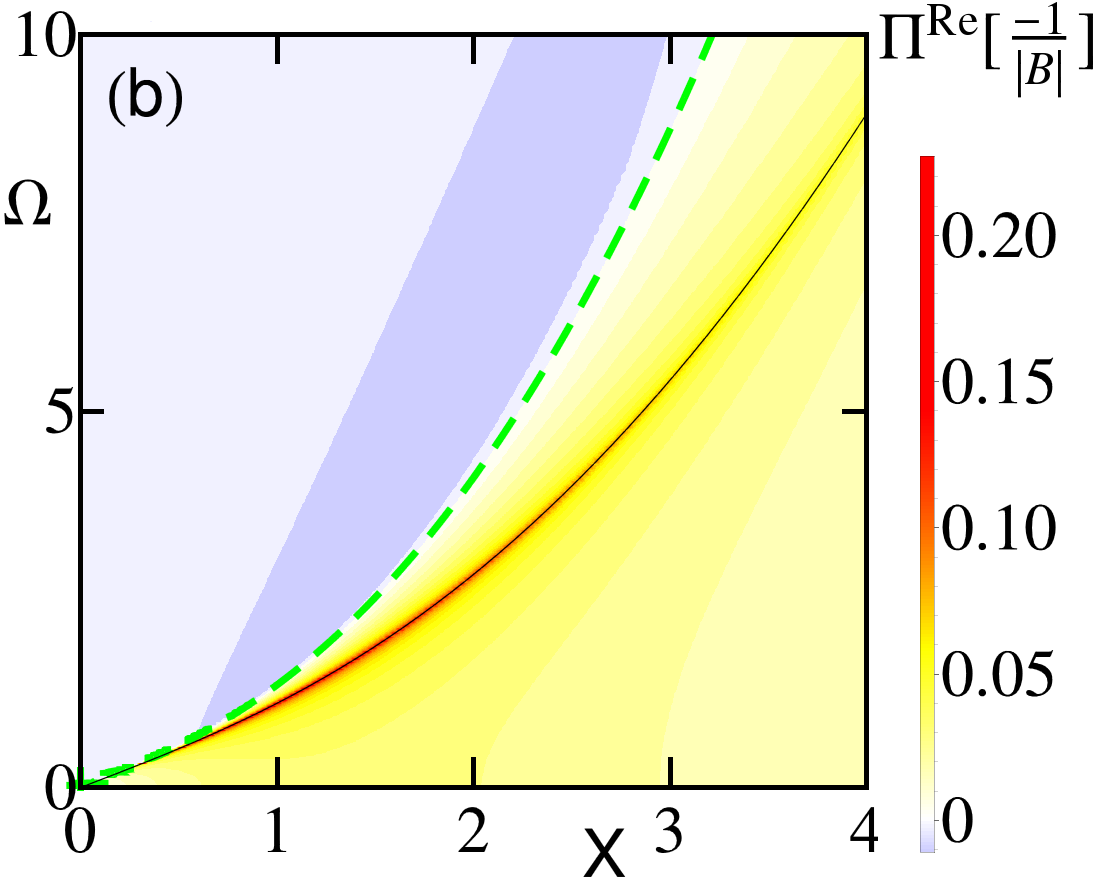}
    \includegraphics[width=4.2cm]{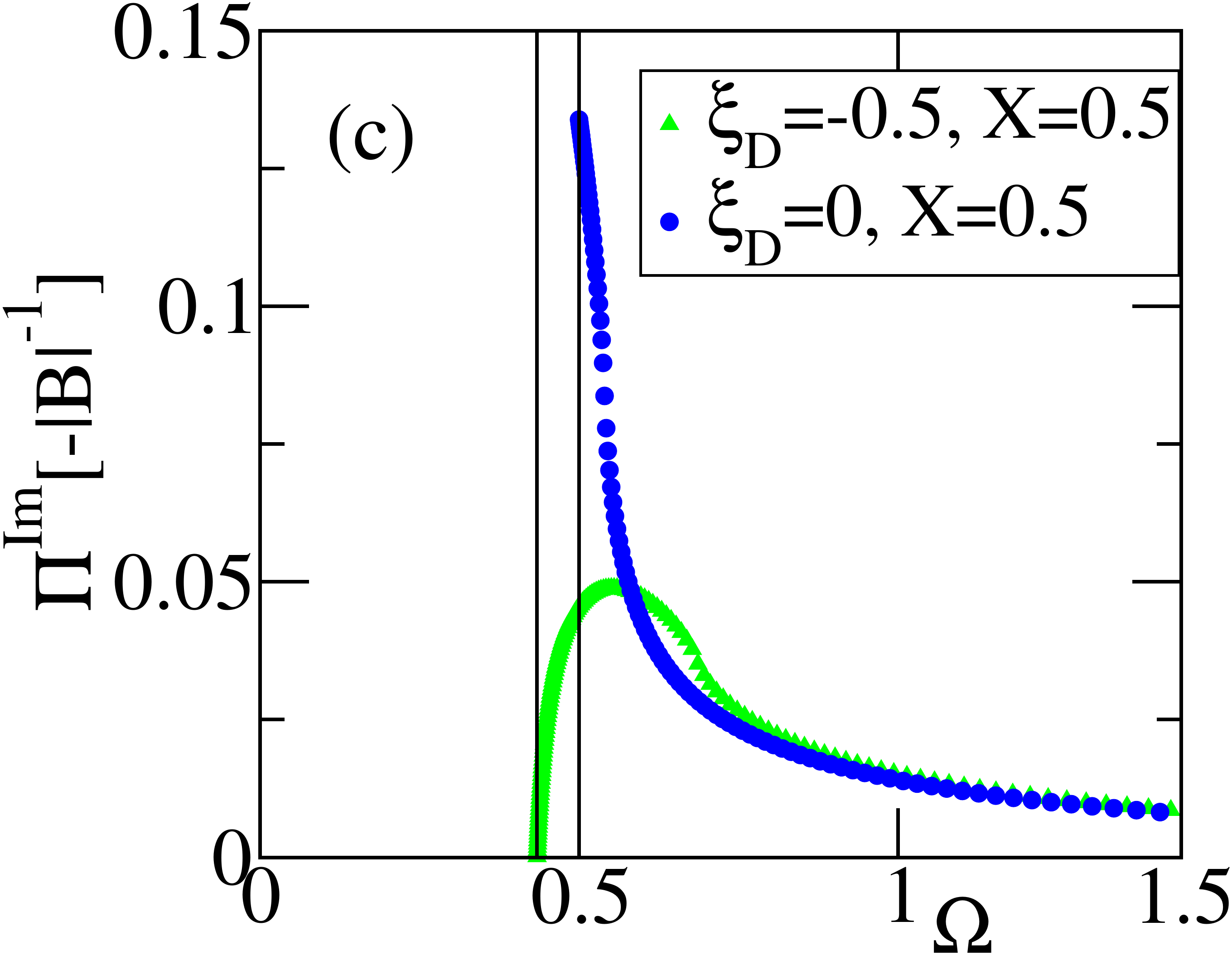}
    \includegraphics[width=4.2cm]{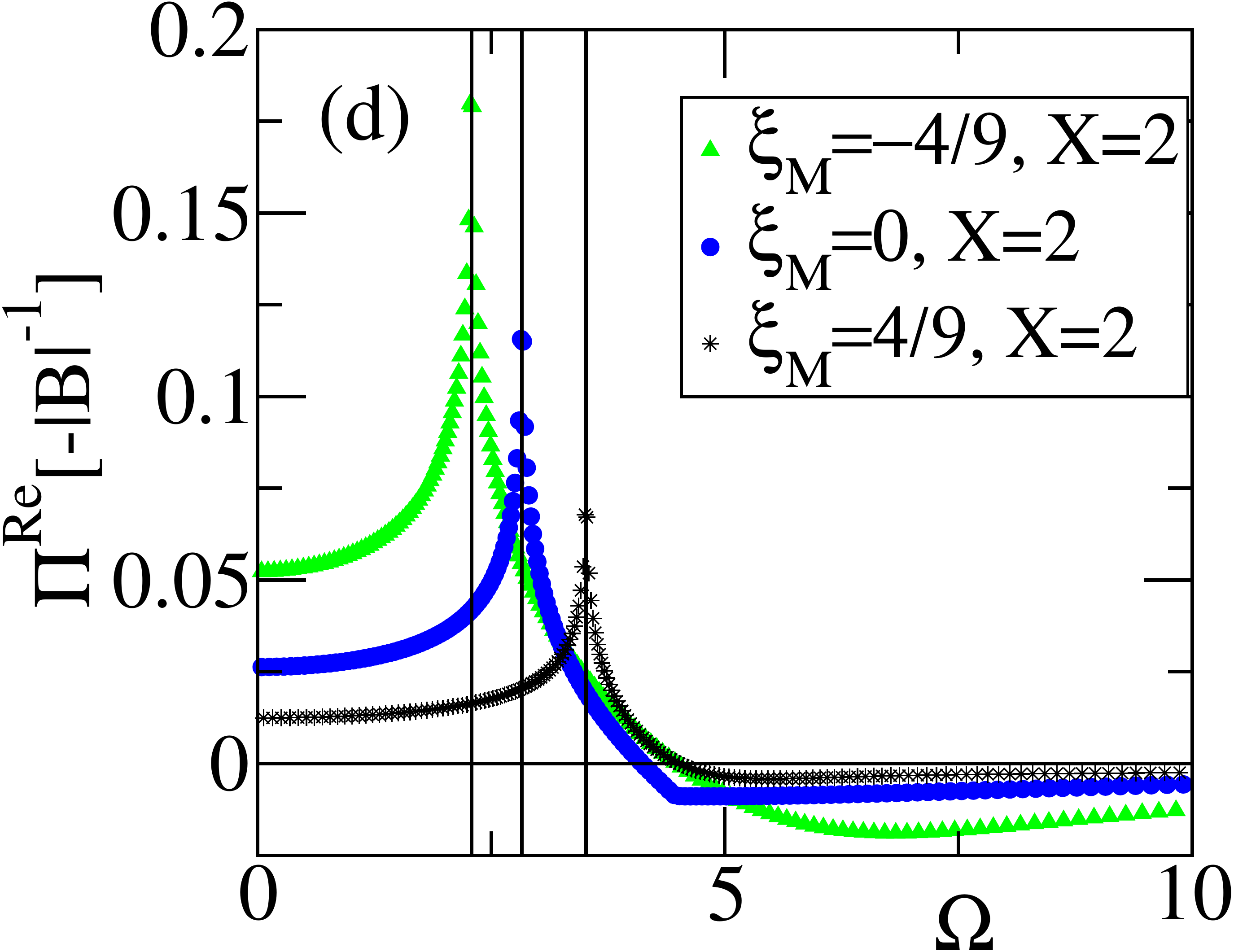}
    \caption{Real (a) and imaginary (b) part of $\Pi^{R}\left(X,\Omega\right)$ of the BHZ model for $\xi_{M}=\xi_{D}=0$.
    In (c) line cuts are drawn of $\Im\left[\Pi^{R}\right]$ for $X=0.5$ with $\xi_{D}=-0.5$, $0$. The line cuts in (d) show $\Re\left[\Pi^{R}\right]$ for 
        $X=2$ with $\xi_{M}=-\frac{4}{9}$, $0$ and $\frac{4}{9}$ with green triangles,
    blue dots and black stars, respectively.
    }
    \label{fig: Polarization}
  \end{figure}

  The imaginary part of the polarization function is calculated from Eq.~($\ref{eq:Pi_par}$), by employing the relation
  $\Im\left[\frac{1}{\omega + \rm i0^{+} }\right]=-\pi\delta\left(\omega\right)$.
  The real part is then obtained by the Kramers-Kronig relation. In Fig.~\ref{fig: Polarization} we plot $\Pi^{R}\left(X,\Omega\right)$ for $\xi_{M}=\xi_{D}=0$.
  The imaginary part in panel (a) is strictly zero below the cut-off frequency $\Omega_{min}$,
  where electronic excitations are forbidden by energy-momentum conservation.
  In the Dirac limit ($X,\Omega\rightarrow0$), the cut-off frequency is $\Omega_{min}=X$,
  where a divergent behavior is observed \cite{gangadharaiah2008}
  because the linear spectrum allows for the existence of a divergent number of particle-hole excitations
  satisfying the energy-momentum conditions.
  In the opposite limit ($X,\Omega\rightarrow\infty$), the polarization function goes
  to zero, due to the vanishing overlap factor, as expected in the 2DEG limit.
  At fixed $X$ and $\Omega\rightarrow\infty$, the imaginary part decays as $\Omega^{-2}$,
  while in a purely Dirac system it shows a $\Omega^{-1}$ decay.
  Whereas in graphene within RPA, the real part of $\Pi^{R}$ is always negative, for the BHZ model, as shown in Fig.~\ref{fig: Polarization}(b),
  $\Re\left[\Pi^{R}\right]$ changes sign and becomes positive in the region above $\Omega_{min}$ indicated by the green dashed line.
  In this anti-screening region, we can search for solutions to Eq.~($\ref{eq:Plasmon Equation}$),
  describing plasmonic resonances.
  Yet, Landau damping of the plasmon mode by single-particle excitation processes can be expected due to the
  finite value of $\Im\left[\Pi^{R}\right]$ in the same region. 

  In Fig.~\ref{fig: Polarization}(c), we plot line cuts of $\Im \left[\Pi^{R} \right]$
  for fixed $X=0.5$ with and without $\xi_{D}$. A finite $\xi_{D}$ strongly changes the polarization for small $X$, as $\Im\left[\Pi^{R}\right]$ goes to zero at $\Omega_{min}$, instead of exhibiting the divergency known from graphene. This is due to the breaking of particle-hole symmetry and
  %, such that at low energy electrons from the conduction band are excited close to the Dirac point, where the density of states goes to zero. Importantly, this reduction of $\Im\left[\Pi^{R}\right]$ 
  greatly diminishes the Landau damping of plasmons. 
  In Fig.~\ref{fig: Polarization}(d), we plot line cuts of $\Re \left[\Pi^{R} \right]$
  for fixed $X=2$ and different Dirac masses.
  One nicely sees that, due to their effect on the overlap factor, a negative (positive) mass
  enhances (diminishes) the features of the polarization function with respect to the $M=0$ case.
  Therefore, in a topologically non-trivial phase the antiscreening ($\Re\left[\Pi^{R}\right]>0$ part)
  effect gets enhanced, which increases the chance of observing plasmons.

  %%%%%%%%%%%%%%%%%%%%%%%%%%%%%%%%%%%%%%%%%%%%%%%%%%%%%%%%%%%%%%%%%%%
  %%%%%%%%%%%%%%%%%%%%%%%%%%%%%%%%%%%%%%%%%%%%%%%%%%%%%%%%%%%%%%%%%%%
  \paragraph{Plasmons.}
  Now, we look for solutions of Eq.~(\ref{eq:Plasmon Equation}) corresponding to plasmon quasi-particles of definite energy and
  momentum (where $\frac{\Gamma}{\Omega} < 1$), likely to be observable in experiments.
    \begin{figure}
    \includegraphics[width=4.2cm]{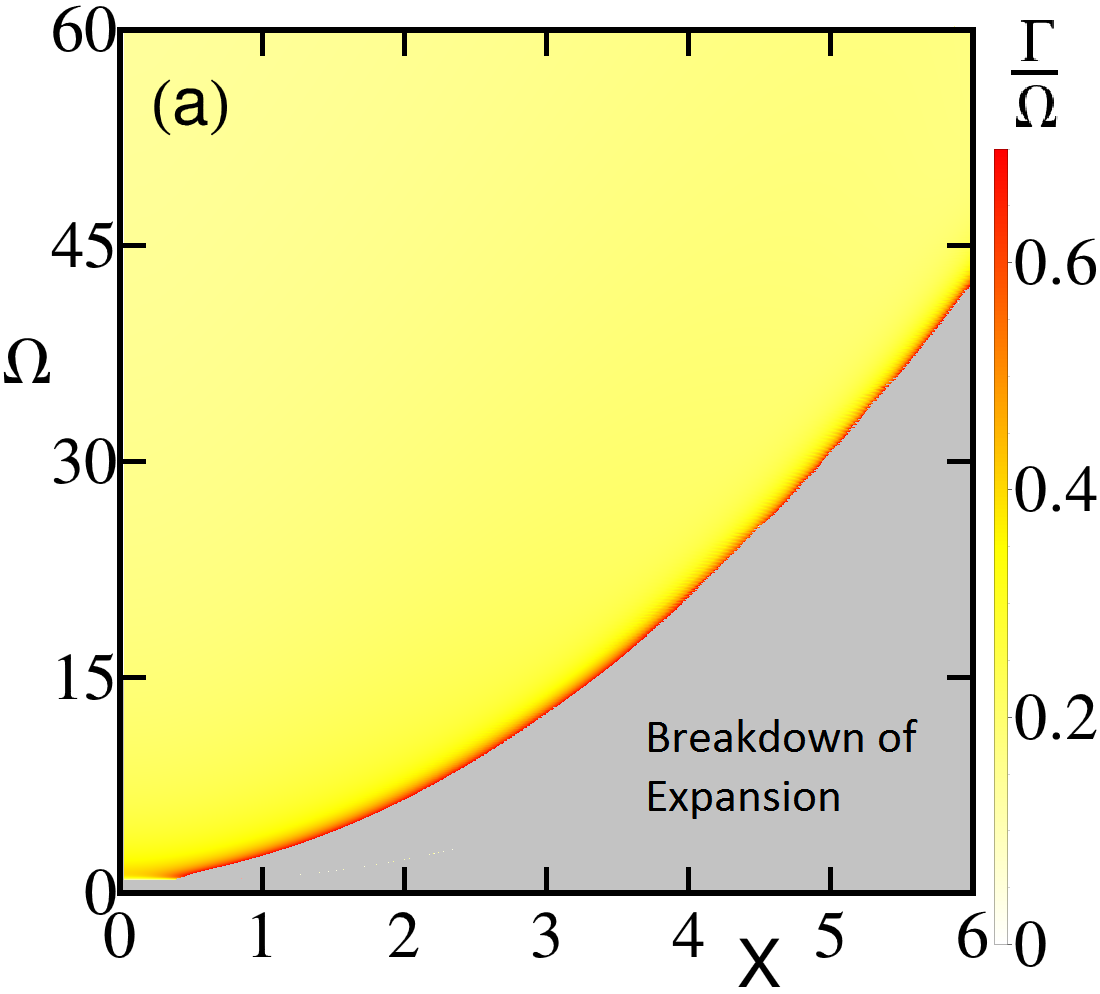}
    \includegraphics[width=4.2cm]{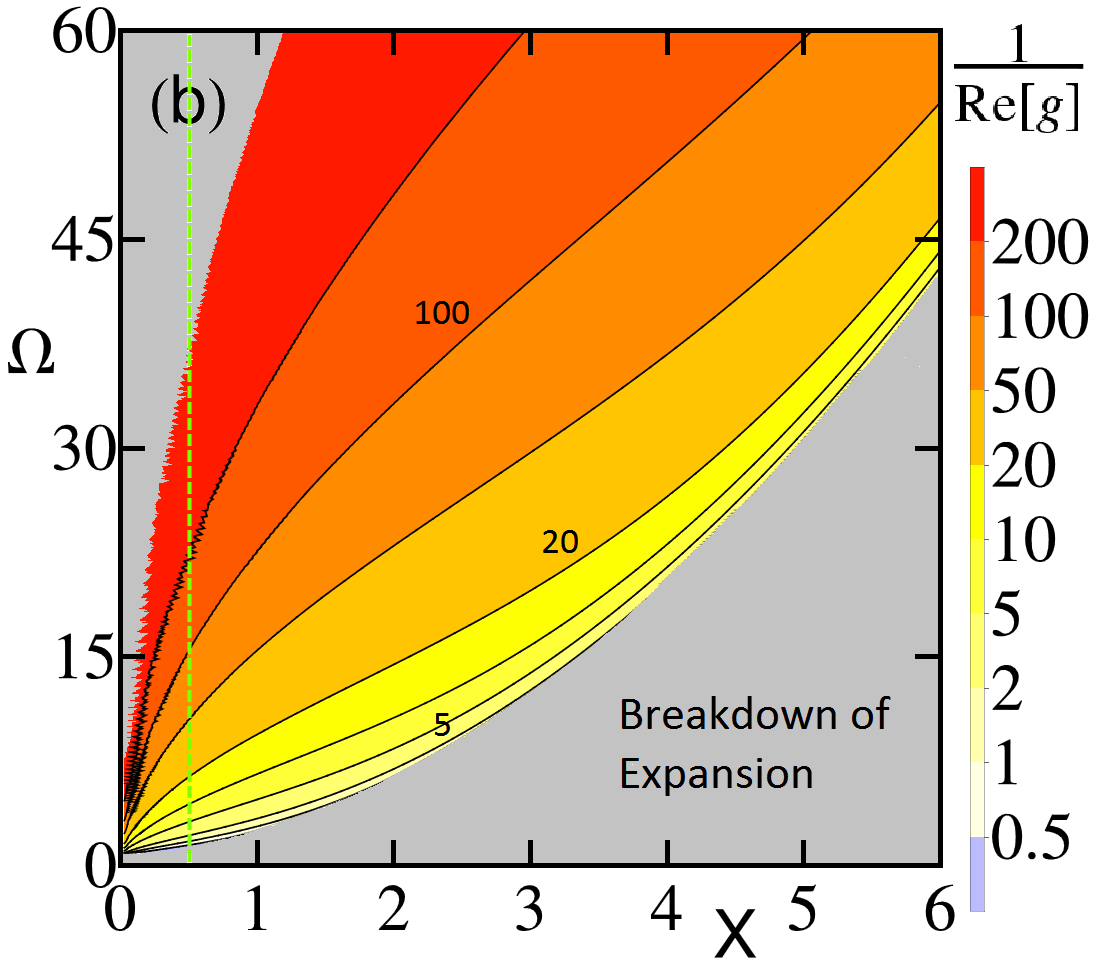}
    \includegraphics[width=4.15cm]{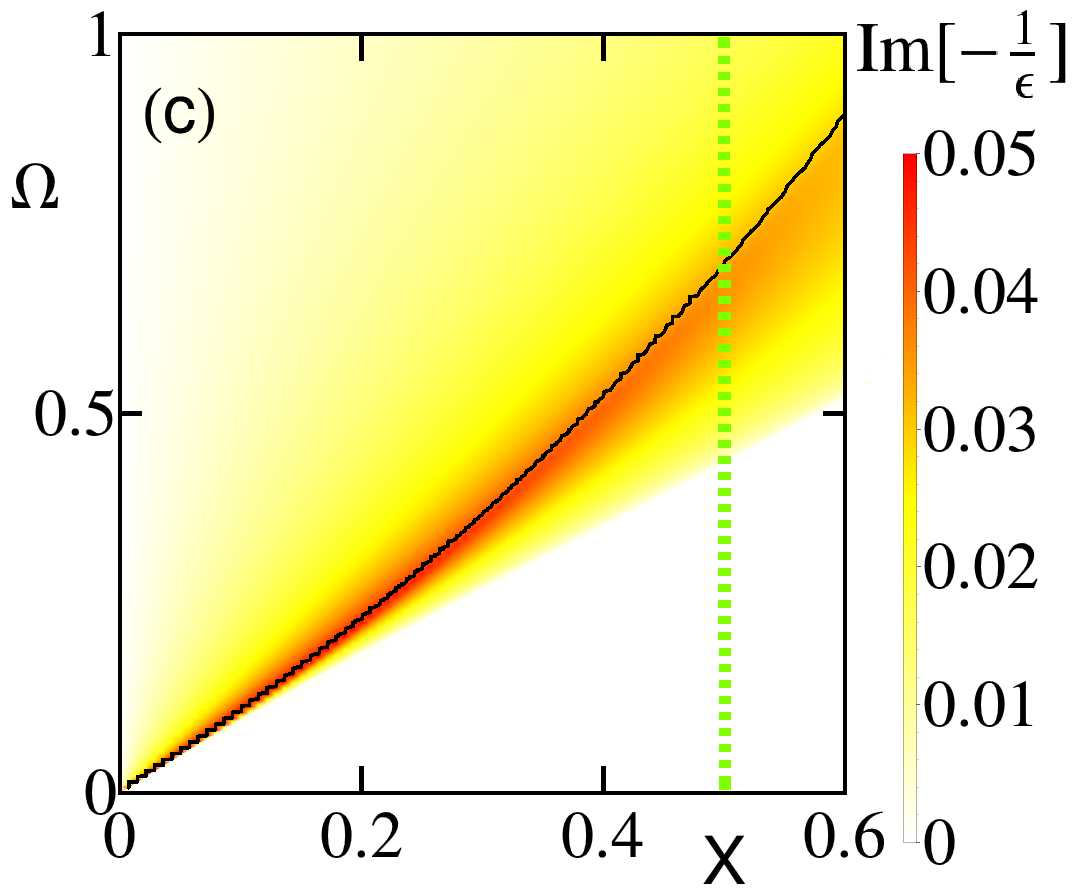}
    \includegraphics[width=4.25cm]{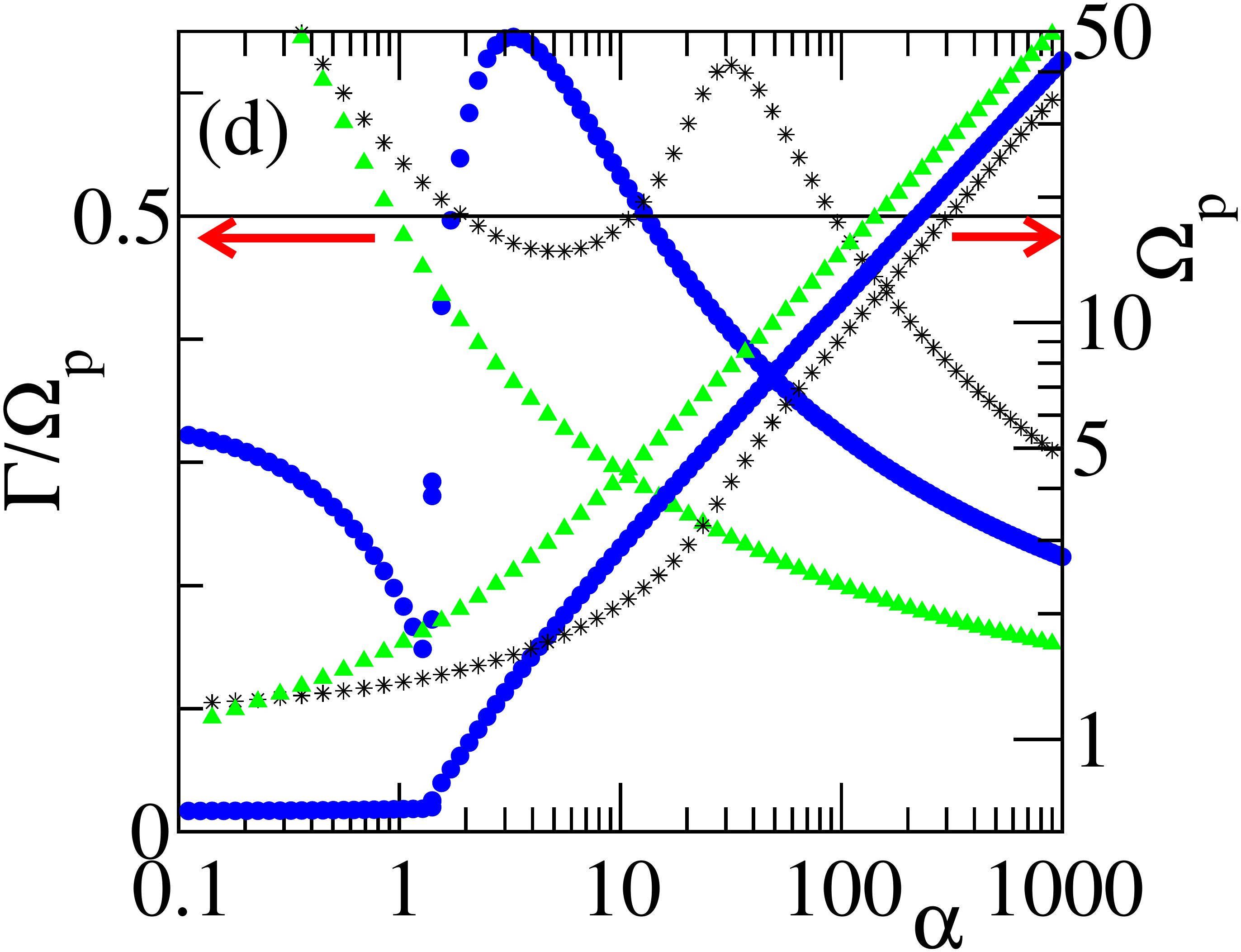}
    \caption{(a) Illustration of the damping ratio $\frac{\Gamma}{\Omega}$ and (b) the function
    $\frac{1}{\Re\left[g\left(X,\Omega-i\Gamma\right)\right]}$, for $\xi_{M}=-\frac{4}{9}$. (c) Plasmon dispersion together
    with the loss function $\Im\left[-\frac{1}{\varepsilon}\right]$ for $\xi_{D}=-0.5$ and $\alpha=0.4$. 
    (d) Plasmon frequency and damping at $X=0.5$ (dashed line in panels (b) and (c)) for $\xi_{M}\in\left\{-\frac{4}{9},\frac{4}{9}\right\}$ and $\xi_{D}=-0.5$ in green triangles,
    black stars and blue dots, respectively. 
    }
    \label{fig: Plasmons}
  \end{figure}
  For this purpose, Eq.~(\ref{eq:Plasmon Equation}) is expanded up to order $(\Gamma/\Omega)^2$~\cite{Fetter, wunsch2007},
  resulting in two equations for real and imaginary part, from which we obtain the plasmon dispersion and its damping factor, respectively.
  We find that the damping ratio $\frac{\Gamma}{\Omega}\left(X,\Omega\right)$
  parametrically depends on $\xi_{M}$ and $\xi_{D}$ only, but not on the interaction strength $\alpha$.
  This behavior implies that Landau damping is due to the
  non-interacting single-particle excitations.
  In Fig.~\ref{fig: Plasmons}(a), we plot the damping ratio $\frac{\Gamma}{\Omega}$ for a TI phase with $\xi_{M}=-\frac{4}{9}$.
  For frequencies $\Omega\gg\Omega_{min}$, the system is sufficiently undamped
  and the expansion in $\Gamma/\Omega$ is justified.
  In Fig.~\ref{fig: Plasmons}(b), we show the contour plot of $\frac{1}{\Re\left[g\left(X,\Omega-i\Gamma\right)\right]}$ for $\xi_{M}=-\frac{4}{9}$.
  The isolines $\frac{1}{\Re\left[g\left(X,\Omega-i\Gamma\right)\right]}=\alpha$ yield the plasmon dispersion curve for different interaction strength $\alpha$.
  The plasmon dispersion relation may vary from the square root dependence on $X$ known from doped graphene~\cite{hwang2007} and ordinary 2DEGs. In our case, we recover this behavior for $\alpha\rightarrow\infty$ but observe instead an almost linear dependence on $X$ (for small $X$) in the limit of $\alpha\rightarrow0$.
  %
 % A similar flattening has been predicted for intrinsic plasmons in graphene at finite temperature~\cite{dassarma2013}.
  Plasmons are also revealed as peaks in the loss function $\Im\left[-\frac{1}{\varepsilon}\right]$.
  In Fig.~\ref{fig: Plasmons}(c) we compare the plasmon
  dispersion calculated from Eq.~(\ref{eq:Plasmon Equation}) with the loss function for
  an interaction strength of $\alpha=0.4$ and $\xi_{D}=-0.5$. 
  Evidently, plasmons are easily resolved and in perfect agreement with our analytical calculation.
  This is the key results of this paper: the interplay between Dirac and Schr{\" o}dinger fermions leads to a plasmonic excitation, which is absent in the limiting cases of a pure Dirac or Schr{\" o}dinger system. 
  %To the best of our knowledge, we predict, for the first time, plasmonic excitations of a 2D electronic system
 % in the intrinsic limit and at zero temperature within RPA.
  %
  
  %Indeed, plasmons have been proposed to exist also in intrinsic graphene but only at finite temperature~\cite{vafek2006,dassarma2013} or
 % when a special kind of higher order diagrams (beyond RPA) are taken into account~\cite{gangadharaiah2008}, which is still in debate~\cite{sodemann2012}. Hence, the physical reason for %the appearance of plasmons in our case is rather different from the previously reported ones:
  %
  %The plasmonic solutions found here are due to the mixed Dirac/Schr{\" o}dinger nature of the electronic system.

  In Fig.~\ref{fig: Plasmons}(d), we plot the plasmon frequency
  and damping as a function of $\alpha$ at fixed $X=0.5$ for a NI and a TI phase ($\xi_{D}=0$) as well as for $\xi_{D}=-0.5$.
  As the plasmon frequency increases with $\alpha$, for $\alpha\rightarrow\infty$ the damping ratio $\frac{\Gamma}{\Omega_p}$ decreases to
  values below $\frac{\Gamma}{\Omega_p}\lesssim 0.2$.
  Notably, at large $\alpha$, the TI phase yields a larger plasmon frequency and is considerably less damped than the NI phase.
  This behavior directly stems from the enhancement of the anti-screening region of the polarization function due to the overlap factor enhancement in the TI phase.
  In the opposite limit, $\alpha\rightarrow0$, the excessive damping leads to a breakdown of our expansion.
  For finite $\xi_{D}$ on the other hand, the plasmon damping has a minimum around $\alpha\approx1$, which prevails to even smaller interaction strengths. This behavior is a direct consequence of the smaller $\Im\left[\Pi^{R}\right]$ leading to reduced Landau damping for small plasmon frequencies. 
  %The latter one has the smallest damping ratio in this regime. Interestingly, in the NI phase, we observe a second low damped region for smaller values of $\alpha$ and
  %$\Omega_p$ [see $\alpha\approx3.5$ and $\Omega\approx2.4$ in Fig.~\ref{fig: Plasmons}(d)] in addition to the one for large
  %$\alpha$ and $\Omega_p$ visible for all three cases.

\paragraph{Experimental realisation.}
  Considering experimental parameters for Hg(Cd)Te structures~\cite{buttner2011,schmidt2009}, one finds roughly $\xi_{D}\leq-0.5$, $q_{0}\approx0.4\ \frac{1}{\mathrm{nm}}$, $E_{0}\approx 140\ \mathrm{meV}$ and masses $M$ with absolute values up to several meV. The interaction strength is about $\alpha\approx0.27$, resulting from $\alpha\approx4/\varepsilon_{r}$ with an average $\varepsilon_{r}=15$ from the CdTe substrate ($\varepsilon_{r}=10$) and HgTe ($\varepsilon_{r}=20$).   
  As recently reported for the surface states of a 3D TI~\cite{pietro2013}, plasmons with a ratio of $\frac{\Gamma}{\Omega_p}=0.5$ are
  still observable in experiments. From Fig.~\ref{fig: Plasmons}(d) we find $\frac{\Gamma}{\Omega_p}\approx0.3$ for $\alpha\approx0.27$ and $\xi_{D}=-0.5$. The wave vector and frequency of the plasmon is extracted from Fig.~\ref{fig: Plasmons}(c) to be $q\in\left[0.1,0.6\right]q_{0}=\left[0.04,0.24\right]\frac{1}{\mathrm{nm}}$ and $\omega\in\left[0.1,0.8\right]\mathrm{\frac{E_{0}}{\hbar}}=\left[21,170\right]\mathrm{THz}$, respectively, where the lower bound stems from the merging of plasmon and single-particle background for $X,\Omega\rightarrow0$. This momentum and frequency range is of the right order for experimental techniques like Raman spectroscopy~\cite{geutrs}. 
A finite temperature in experiments can lead to doping by thermal excitations. At the temperature of liquid helium, one finds $k_{B}T_{He}\approx0.35\ \mathrm{meV}$ with $k_{B}$ the Boltzmann constant. Thus, the plasmons resulting from thermal excitations occur on an energy and momentum scale at least two orders of magnitude smaller than the plasmons discussed in this paper, making it possible to fully separate them or to suppress them with a small gap $k_{B}T<|M|$. 
We conclude that the plasmonic resonances discussed above are measureable, e.g. with Raman spectroscopy on Hg(Cd)Te quantum wells. 
%%%%%%%%%%%%%%%%%%%%%%%%%%%%%%%%%%%%%%%%%%%%%%%%%%%%%%%%%%%%%%%%%%%%
%%%%%%%%%%%%%%%%%%%%%%%%%%%%%%%%%%%%%%%%%%%%%%%%%%%%%%%%%%%%%%%%%%%%
  \paragraph{Optical conductivity.}
  From the knowledge of the polarization function, we can calculate the bulk optical conductivity of the system,
  defined by \cite{stern1967}
  \begin{equation}
    \sigma_{0}\left(\Omega\right)=\underset{X\rightarrow0}{\lim}\frac{\Omega}{X^{2}}\left|B\right|\Pi^{R}\left(X,\Omega\right).
  \end{equation}
  Notably, $\sigma_{0}$ is a universal function depending only on $\Omega$ and $\xi_{M}$.
  An analytical calculation yields
  \begin{equation*}
    \Im\hspace{-0.02cm}\left[\hspace{-0.02cm}\sigma_{0}\hspace{-0.02cm}\left(\hspace{-0.02cm}\Omega\hspace{-0.02cm}\right)\hspace{-0.02cm}\right]
    \hspace{-0.05cm}=\hspace{-0.05cm}
    -\hspace{-0.03cm} \left[
    \hspace{-0.06cm}\frac{1}{W}\hspace{-0.05cm}
    +\hspace{-0.05cm} \frac{1\hspace{-0.04cm}+\hspace{-0.04cm}4\xi_{M}}{\Omega^{2}} \hspace{-0.1cm}
    \left(\hspace{-0.03cm}\frac{1\hspace{-0.04cm}+\hspace{-0.04cm}2\xi_{M}}{W}\hspace{-0.05cm}
    -\hspace{-0.05cm}\frac{1}{2}\hspace{-0.06cm}\right)
     \hspace{-0.1cm}\right]
    \Theta\hspace{-0.05cm}\left(\hspace{-0.02cm}\Omega\hspace{-0.06cm}-\hspace{-0.06cm}2
    \hspace{-0.03cm}\left|\hspace{-0.02cm}\xi_{M}\hspace{-0.02cm}\right|\hspace{-0.04cm}\right),
    \label{eq:Optical conductivity}
   \end{equation*}
  where $W=2\sqrt{1+4\xi_{M}+\Omega^{2}}$.

\begin{figure}
    \includegraphics[width=5.5cm]{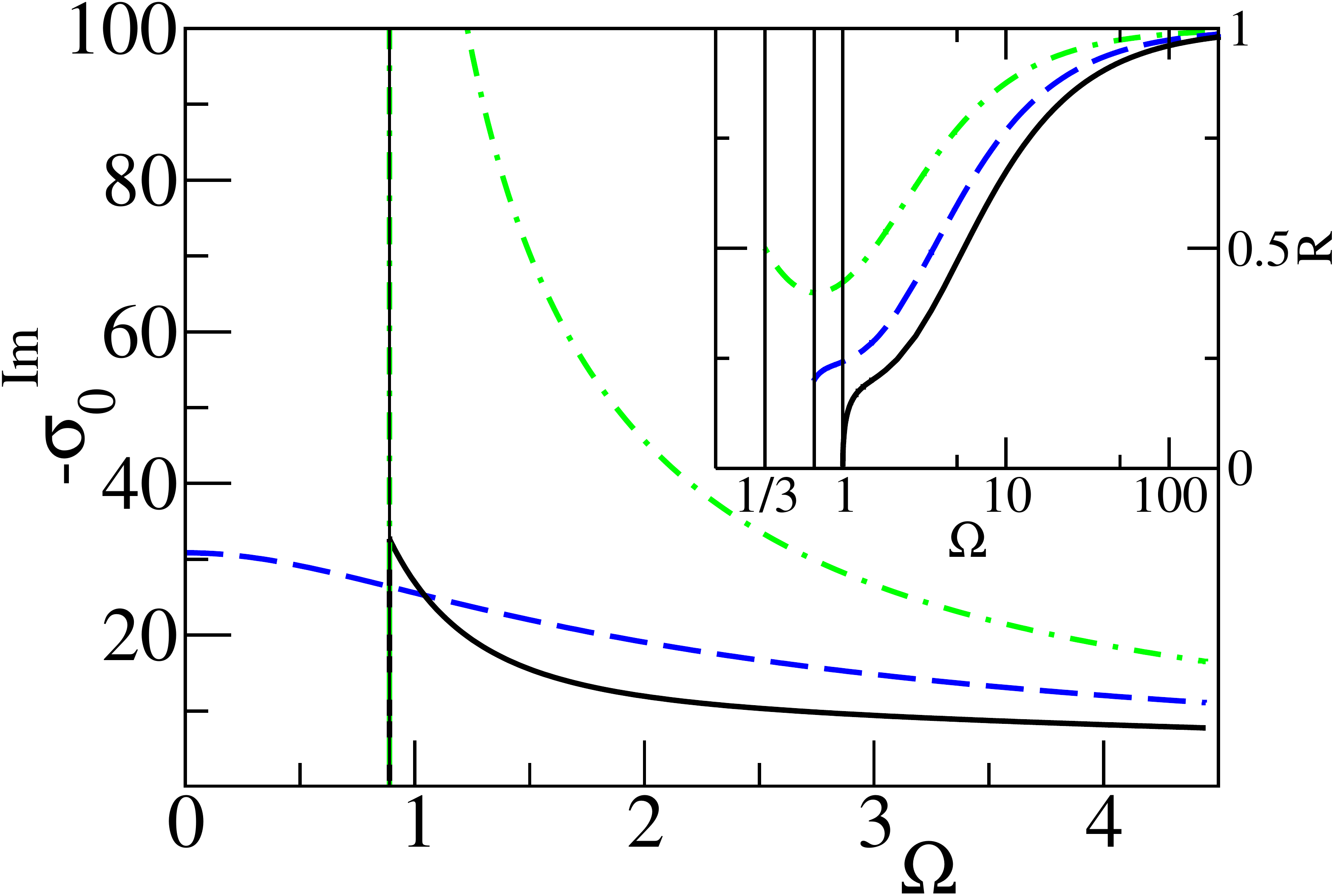}
    \caption{Illustration of $\Im\left[\sigma_{0}\left(\Omega\right)\right]$ for $\xi_{M}=-\frac{4}{9}$, $0$ and $\frac{4}{9}$ with
    dot-dashes, long dashes and a solid line, respectively.
    The inset shows the ratio $R=\frac{\Im\left[\sigma_{0}\left(\Omega\right)\right]_{M>0}}{\Im\left[\sigma_{0}\left(\Omega\right)\right]_{M<0}}$
    for $\left|\xi_{M}\right|=\frac{1}{6}$, $\frac{1}{3}$ and $\frac{1}{2}$ in dot-dashes, long dashes and a solid line, respectively.
    }
    \label{fig: Pi q=00003D0}
  \end{figure}

  In Fig.~\ref{fig: Pi q=00003D0},
  we plot $\Im\left[\sigma_{0}\left(\Omega\right)\right]$ for the different masses $\xi_{M}=-\frac{4}{9}$, $0$ and $\frac{4}{9}$.
  Compared to the massless case, both positive and negative Dirac mass lead to a peak just above $\Omega_ {min}=|2\xi_M|$.
  The signal from the TI phase is much stronger, diverging for
  $\left|\xi_{M}\right|\rightarrow\frac{1}{2}$,  which is the threshold of turning the band structure into a Mexican hat shape.
  Such difference between the trivial and topological phases is emphasized in the inset of Fig.~\ref{fig: Pi q=00003D0}, where we plot the ratio
  $R=\frac{\Im\left[\sigma_{0}\left(\Omega\right)\right]_{M>0}}{\Im\left[\sigma_{0}\left(\Omega\right)\right]_{M<0}}$.
  This behavior can be explained by considering the combined effects of the overlap factor and the phase space for the excitation process.
  In the TI phase of the BHZ model, the conduction and valence bands flatten for $X<|\xi_M|$, with respect to the NI phase.
  This fact enormously increases the number of finite momentum states, available for an
  excitation just above $\Omega_{min}=|2\xi_M|$.
  The optical conductivity shows also a qualitatively different behavior as a function of $\left|\xi_{M}\right|$.
  Indeed, $\Im\left[\sigma_{0}\left(\Omega\right)\right]$ increases (decreases) with
  increasing $\left|\xi_{M}\right|$ for a negative (positive) mass.

To conclude, we have analyzed the influence of Coulomb interaction -- within RPA -- on a mixed Dirac/Schr{\" o}dinger fermion system described by the BHZ model. This model describes well a 2D TI realized in Hg(Cd)Te quantum wells. In the intrinsic limit, we could find observable plasmon solutions which is a remarkable consequence of the peculiar electronic spectrum of the model. These plasmons occur for parameters suitable for experiments (like Raman spectroscopy) on Hg(Cd)Te quantum wells. Furthermore, we have predicted that a measurement of the optical conductivity at finite frequency (to be precise, above $\Omega=|2\xi_M|$) yields a direct way to distinguish between the two topological phases of Hg(Cd)Te quantum wells.
  %%%%%%%%%%%%%%%%%%%%%%%%%%%%%%%%%%%%%%%%%%%%%%%%%%%%%%%%%%%%%%%%%%%%
  %%%%%%%%%%%%%%%%%%%%%%%%%%%%%%%%%%%%%%%%%%%%%%%%%%%%%%%%%%%%%%%%%%%%

  We acknowledge interesting discussions with M. Polini and financial support by the DFG (SPP1666 and the DFG-JST research unit {\it Topotronics}) as well as the Helmholtz Foundation (VITI).

  %%%%%%%%%%%%%%%%%%%%%%%%%%%%%%%%%%%%%%%%%%%%%%%%%%%%%%%%%%%%%%%%%%%%
  %%%%%%%%%%%%%%%%%%%%%%%%%%%%%%%%%%%%%%%%%%%%%%%%%%%%%%%%%%%%%%%%%%%%

%  \begin{acknowledgements}
%  \end{acknowledgements}

%%%%%%%%%%%%%%%%%%%%%%%%%%%%%%%%%%%%%%%%%%%%%%%%%%%%%%%%%%%%%%%%%%%%
%%%%%%%%%%%%%%%%%%%%%%%%%%%%%%%%%%%%%%%%%%%%%%%%%%%%%%%%%%%%%%%%%%%%

% \bibliographystyle{IEEEtran}
%\bibliography{IEEEabrv,bibf}
%\bibliography{TopIns_lett}

\begin{thebibliography}{24}
\bibitem{pseudo} In order to have a band structure with a minimum at the $\Gamma$-point, we restrict ourselves to masses $M>-\frac{1}{2}\frac{A^{2}}{|B|}$.
\end{thebibliography}

\end{document}